\documentstyle[preprint,prl,aps]{revtex}

\begin{document}
\draft
\centerline{\bf Pattern Competition in the Photorefractive Semiconductors}
\vskip 10 mm
\centerline{Yuo-Hsien Shiau$^{1,*}$ and Chin-Kun Hu$^{2,3,+}$}
\centerline{$^1$
National Center for Theoretical Sciences (Physics Division) 101,
Section 2 Kuang Fu Road,}
\centerline{
Hsinchu, Taiwan 300, ROC}
\centerline{$^2$
Institute of Physics, Academia Sinica, Taipei 11529, Taiwan}
\centerline{$^3$Department of Physics, National Dong Hwa University, 
Hualien 97401, Taiwan}

\begin{abstract}
We analytically study 
the photorefractive Gunn effect in n-GaAs
subjected to two external laser beams which form
a moving interference pattern (MIP) in the semiconductor.
When the intensity of the spatially independent part of the MIP,
denoted by $I_0$, is small, the system has a
periodic domain train (PDT), consistent with the results
of linear stability analysis.
When $I_0$ is large, the space-charge field induced by the MIP
will compete with the PDT and 
result in complex dynamics, including driven chaos
via quasiperiodic route.
\end{abstract}

\vskip 0.5truecm
\noindent{PACS number(s): 72.20.Ht, 42.70.Nq, 05.45.-a, 05.45.Yv}

\vskip 1 cm
\noindent{Keywords: A. Semiconductors; D. Pattern Competition; E. Nonlinear Analysis}

\narrowtext
\newpage

Spatiotemporal behavior of Gunn domains \cite{Rid,Gunn}
in semiconductors
is an interesting and important current research problem
\cite{Seg,Sub,Bon,s3,s1,s2}.
Gunn diodes are not only good systems for studying nonlinear dynamics,
but also potentially important for various 
applications \cite{s1,s2,ye}. In 1996-1997
Segev {\em et al.} \cite{Seg} and Suba\v cius {\em et al.} \cite{Sub}
showed, respectively, that optical waves can excite
multiple Gunn-domain formation with spatial periodicity,
called periodic domain trains (PDT),
in deep-impurity doped GaAs and
semi-insulating GaAs; both groups used the well
known Kroemer's criterion \cite{Rid} to determine the number of
Gunn domains.
In 1998, Bonilla {\em et al.} \cite{Bon} numerically
studied the same dynamical system as Segev {\em et al.}; however,
their findings are mostly inconsistent with the predictions
of Segev {\em et al.}. Bonilla {\em et al.} \cite{Bon} found
that the number of Gunn domains in the system is not consistent with
the prediction of Segev {\em et al.} and in many cases the system
shows complex or chaotic behavior.
Very recently,
Shiau {\em et al.} used a linear stability analysis (LSA) to study a
shallow-impurity doped GaAs with length $L$ subjected
to two external laser beams which form a moving interference pattern
(MIP) of intensity $I(x,t)=I_0[1+m\;\mbox{cos}(Kx+\Omega t)]$
and predicted PDT in the Gunn diode \cite{s1}.
Here $\Omega $ is the frequency difference of the two optical waves,
$K=2\pi /\Lambda $ is the interference wave number, $\Lambda $ is the 
grating period and is much smaller than $L$,
$m$ is the modulation depth, and $I_0$ is the average intensity. 
With consideration of time-delay feedback, 
the diode will emit high-dimensional chaotic 
microwaves, which is potentially useful for secure microwave communications,
memory devices, applications involving photorefractive effects, 
{\em etc.} \cite{s1}.

The purpose of this Communication is to 
clarify different findings of \cite{Seg}, \cite{Sub}, and \cite{Bon}
using a nonlinear analytic  method to
study the system of \cite{s1,s2} as $L \to \infty$.
In this approach, the intensity of the spatially independent part of 
the MIP, $I_0$, plays an important
role in determining the simple dynamics (i.e., PDT) or complex dynamics of Gunn domains.
When $I_0$ is small enough, the system shows simple
dynamics consistent with the result of the LSA,
and Kroemer's criterion is still valid.
When $I_0$ is large, the space-charge field induced by the MIP
will compete with the PDT and result in complex dynamics. 
This picture is also useful for understanding
the results of other groups \cite{Seg,Sub,Bon} discussed above.

We consider two optical waves
incident on a $n$-GaAs with a biased voltage $V$.
The photon energy is just
above the bandgap of GaAs, i.e., 1.42 eV, so that electron-hole pairs can
be generated by optical excitation. 
The generation-recombination processes include
complete thermal ionization of the donor, generation of electron-hole
pairs by the MIP with rate $g$, and recombination of 
electron-hole pairs with rate $\gamma$.   
Therefore we can write
the dynamical equations for the electric field $E$, the free
electron density $n$, and the free hole density $p$ as:
\begin{equation}
\frac{\partial E}{\partial x}=\frac e\epsilon (n-N_D^{*}-p),  \label{2}
\end{equation}
\begin{equation}
\frac{\partial n}{\partial t}=gI(x,t)-\gamma np-\frac \partial {\partial x}%
\left[ nv(E)-D_n\frac{\partial n}{\partial x}\right] ,	\label{3}
\end{equation}
\begin{equation}
\frac{\partial p}{\partial t}=gI(x,t)-\gamma np+\frac \partial {\partial x}%
\left[ p\mu _pE+D_p\frac{\partial p}{\partial x}\right] ,  \label{4}
\end{equation}
\begin{equation}
V=\int_0^LE(x,t)dx.  \label{5}
\end{equation}
Eq. (\ref{2}) is for the Gauss law. Eq. (\ref{3}) and Eq. (\ref{4})
represent the continuity equations for the electrons and holes, respectively.
The circuit condition is given at Eq. (\ref{5}).
Here $v(E)$ displays an N-shaped negative differential mobility (NDM).
The homogeneous solution of Eqs. (\ref{2})-(\ref{5}) is $E_0=V/L$, 
$n_0=(N_D^*+\sqrt {N_D^{*2}+4gI_0/\gamma})/2$, 
and $p_0=(-N_D^*+\sqrt {N_D^{*2}+4gI_0/\gamma})/2$. 
When $I_0$ is very small, $n_0\approx N_D^*+gI_0/N_D^*\gamma$, 
$p_0\approx gI_0/N_D^*\gamma$, and a natural expansion parameter for
high order corrections is
 $\bar\epsilon =p_0/{N_D^*}= gI_0/{N_D^*}^2\gamma$.
We can rearrange the model equations
by the following rescaling process for the dynamical variables 
and parameters in Eqs. (\ref{2})-(\ref{5}):
$E\rightarrow E, \hspace{.2in} n\rightarrow n, \hspace{.2in}
p\rightarrow \bar\epsilon p, \hspace{.2in} I\rightarrow \bar\epsilon I.$
The small parameter $\bar\epsilon$ can
distinguish the large variables (i.e., $E$ and $n$) from small variable
$p$ in our model. Rescaling variables in
Eqs. (\ref{2})-(\ref{5}), we have \cite{s2}
${\partial E}/{\partial x}= e (n-N_D^*)/\epsilon
+O (\bar\epsilon)$,
$ {\partial n}/{\partial t}=- {\partial}
[nv(E)-D_n {\partial n}/{\partial x}]/{\partial x}
+O(\bar\epsilon)$.
These results and  Eq. (\ref{4}) implies 
\begin{equation}
\frac{\partial E}{\partial t}=-\frac 1{\epsilon}eN_D^*{v}(E)
   -{v}(E)\frac{\partial E}{\partial x}+{D_n}\frac
{\partial^2E}{\partial x^2}+\frac 1{\epsilon}J_{tot}(t),
\label{8}
\end{equation}

\begin{equation}
\frac {\partial p}{\partial t}=gI(x,t)-
\gamma (N_D^*+\frac\epsilon e\frac {\partial E}{\partial x})p
+\frac {\partial}{\partial x}
\left(p\mu_pE+D_p\frac {\partial p}{\partial x}\right),
\label{9}
\end{equation}
where $J_{tot}(t)$ is time-dependent current density.
The physical meaning of Eq. (\ref{8}) is quite clear, which 
represents the Gunn-domain formation and with drift velocity $v(E_0)$
\cite{Rid}. Then, Eq. (\ref{8}) confirms our LSA results 
in Ref. \cite{s1}. In the following we want to show the detailed structure
of PDT.

In order to get the solution of Eq. (\ref{8}), 
an ansatz of the electric field, we set
$E(x,t)= E_0+[E_s(x,t)e^{i\bar {K}(x-v_0 t)}
+E_s^*(x,t)e^{-i\bar {K}(x-v_0 t)}]/2$,
where $v_0=v(E_0)$.
$E_s(x,t)$ is a new variable corresponding to the
amplitude of electric-field domains and $E_s^*(x,t)$ is its complex
conjugate; $\bar K$ ($=2\pi/L$) is a bulk property of semiconductor.
Since $\bar K$ is small, the spatiotemporal
behavior of the dynamical system within the linear dimension
$\Lambda$ is dominated by the behavior of $E_s(x,t)$.
Actually, $\bar K$ is an irrelevent parameter which is 
verified by the numerical works of \cite{Sub}. 
When $E_0\gg |E_s|$, we expand
$v(E)$ as:
$v(E)\approx
v_0+v_0^{(1)}(E-E_0)+{v_0^{(2)}}(E-E_0)^2/2
+{v_0^{(3)}}(E-E_0)^3/6$,
where $v_0^{(j)}=d^jv(E)/dE^j|_{E=E_0}$.
Equation (\ref{8}) and above expansions for $E(x,t)$ and $v(E)$ imply
\begin{eqnarray}
\frac {\partial {E_s}}{\partial t}&=&
      (-\frac e\epsilon N_D^*v_0^{(1)}-D_n{\bar K}^2)E_s
      +(i2D_n\bar K-v_0)\frac {\partial {E_s}}{\partial x}
      -\frac 18v_0^{(2)}
       \left(E_s^2\frac {\partial {E_s^*}}{\partial x}+2|E_s|^2\frac
      {\partial {E_s}}{\partial x}\right)\nonumber\\
      & &+D_n\frac {\partial^2 E_s}{\partial x^2}
      -(\frac 18\frac e\epsilon N_D^*v_0^{(3)}+
      i\frac 18v_0^{(2)}\bar K){|E_s|}^2E_s.
\label{12}
\end{eqnarray}
With $\bar K\rightarrow$ 0 (as length $L \to \infty$) and
$E_s\sim \Psi (t)e^{iKx}$, 
Eq. (\ref{12}) implies 
\begin{equation}
\frac {d\Psi}{dt}= -C\Psi
      -\frac 18\left(\frac e\epsilon N_D^*v_0^{(3)}+iKv_0^{(2)}\right )
      {|\Psi|}^2\Psi.
\label{13}
\end{equation}
Here $C=(e N_D^*v_0^{(1)}/\epsilon+D_nK^2+iKv_0)$.
With $\Psi (t)=R(t)$exp$i\Theta (t)$, Eq. (\ref{13}) implies 
$ R(t)=-(a/b)^{1/2}[{r_0^2e^{-2at}}/
         (r_0^2(1-e^{-2at})+a)]^{1/2}$,
$ \Theta (t)=-Kv_0t-\frac 18Kv_0^{(2)}\int^{t}R^2(t')dt'$,
where $a=eN_D^*v_0^{(1)}/\epsilon+D_nK^2$, 
$b=eN_D^*v_0^{(3)}/8\epsilon$, 
and $r_0$ is related to initial value of $R(t)$.
When $a>0$ (i.e., 
the operating point is within the regime of positive differential mobility PDM), 
$b>0$, and $t\rightarrow \infty$,
$R(t)$ and $\Theta (t)$ tend to $0$ and $-Kv_0t$, respectively. 
These results mean that the origin on the 
complex plane of $\Psi (t)$ is a stable focus, 
which also implies electric-field profile in material is homogeneous, 
i.e., $E(x,t)=E_0$.
When $a<0$ (i.e., the NDM regime), $b>0$,
and $t\rightarrow \infty$,
$R(t)= (-a/b)^{1/2}$ and $\Theta (t)=-K(v_0+ a v_0^{(2)}/8b)t$.
This solution describes a circular trajectory 
with counterclockwise angular frequency 
$K(v_0+ a v_0^{(2)}/8b)$,
radius $(-a/b)^{1/2}$, and transient response time $|a|^{-1}$.
The transition
from stable focus to stable limit cycle is a
supercritical Hopf bifurcation. 
Since $\bar K\rightarrow$ 0,  
$E_s \sim \Psi (t)e^{iKx}$ and $\Psi (t)=R(t)$exp$i\Theta (t)$  imply
for $a < 0$, $b > 0$, and $t\rightarrow \infty$: 
\begin{equation}
E(x,t)=E_0+({-a}/{b})^{1/2}\cos[K(x-v_0t-av_0^{(2)}t/8b)],
\label{15a}
\end{equation}
which means that the PDT
has a spatial period $\Lambda$ and a traveling
velocity $v_0+av_0^{(2)}/8b$.  
Since $v_0$ ($\sim 10^7$ cm/s) is much
larger than $|av_0^{(2)}/8b|$, so we neglect
the term $av_0^{(2)}/8b$ below.

In general, $E_s$ may be written as
\begin{equation}
E_s = \sum_{q=1}^{\infty} \Psi_q (t)e^{iqKx}  
    = \sum_{q=1}^{\infty} -\left(\frac {a_q}b\right)^{1/2}
        \left[\frac {r_0^2e^{-2{a_q}t}}
         {r_0^2(1-e^{-2{a_q}t})+a_q}\right]^{1/2}e^{iqK(x-v_0t)}.
\label{16}
\end{equation}
where $a_q=eN_D^*v_0^{(1)}/\epsilon+D_nK^2q^2$. 
However, Eq. (\ref{16}) needs to be corrected
when we consider the coupling between spatial modes.
This is very important for self-organizing systems.
For example, consider two spatial modes
$\Psi_1 e^{ikx}$ and $\Psi_2 e^{i2kx}$ with $a_2>0>a_1$ and $b>0$.
From equations for $R(t)$ and $\Theta (t)$ , we know that 
$|\Psi_1|\rightarrow (-a_1/b)^{1/2}$
and $|\Psi_2|\rightarrow 0$ when $t \to \infty$. 
Then $|\Psi_2|/|\Psi_1|$ will become zero.
This result implies that only unstable spatial modes (i.e., $a_q<0$)
have contributions for pattern-forming systems. However, if we combine
these two modes, i.e., 
$E(x,t)=E_0+ \sum_{q=1}^2[\Psi_q (t) e^{iqKx}+
\Psi_q^* (t) e^{-iqKx}]/2$, 
and substitute it into Eq. (\ref{8}),
we have  
\begin{equation}
\frac {d\Psi_1}{dt}
=-(a_1+iKv_0)\Psi_1
	       -\sum_{m_1+m_2+\cdots+m_s=1} \frac {\beta_{s,1}}{s!}
		   \Psi_{m_1}\Psi_{m_2}\cdots \Psi_{m_s},
\label{17}
\end{equation}
\begin{equation}
\frac {d\Psi_2}{dt}
=-(a_2+i2Kv_0)\Psi_2
	       -\sum_{m_1+m_2+\cdots+m_s=2} \frac {\beta_{s,2}}{s!}
		   \Psi_{m_1}\Psi_{m_2}\cdots \Psi_{m_s},
\label{18}
\end{equation}
where $m_s=\pm 1$ or $\pm 2$,
$
\beta_{s,m}=eN_D^*v_0^{(s)}/\epsilon
+imsv_0^{(s-1)}K
$,
and $\Psi_q^*=\Psi_{-q}$ \cite{s3}.
Based on the slaving principle \cite{Hak}, 
we know that the stable mode $\Psi_2$ will 
be determined by the unstable mode $\Psi_1$,
and that $\Psi_2$, and hence the ratio
$|\Psi_2| / |\Psi_1| \equiv r_{2,1}(K,N_D^*)$, will be nonzero.
For simplicity, in the following we assume that $r_{q+1,q}$ ($\equiv
|\Psi_{q+1}|/|\Psi_q|$) is independent of $q$ and denote the number  
by $r$, which can be determined by experiments (see below). 
From $E(x,t)=E_0+[E_s(x,t)+E_s^*(x,t)]/2$ and Eq. (\ref{16}), 
for large $t$ we get
\begin{equation}
E(x,t)=E_0+
\left(\frac {-a_1}b\right )^{1/2}
\frac {-r+\cos [K(x-v_0t)]}{1+r^2-2r\cos [K(x-v_0t)]}.
\label{19}
\end{equation}
Therefore, the maximum and minimum fields
are equal to $E_{max}=E_0+(-a_1/b)^{1/2}(1-r)^{-1}$ and
$E_{min}=E_0-(-a_1/b)^{1/2}(1+r)^{-1}$, respectively. 
According to Pockel's effect, the difference in 
refractive index between $E_{max}$ and $E_{min}$ is
$   
-\bar n^3\gamma_{41}(-a_1/b)^{1/2}(1-r^2)^{-1}
$,
where $\bar n$ is refractive index of pure material 
and $\gamma_{41}$ is the electro-optic coefficient
($\bar n=3.3$ and $\gamma_{41}=1.43\times 10^{-10}$ cm/V
for GaAs when the wavelength of the incident 
light is 1 $\mu$m). The profile
of Eq. (\ref{19}), shown in Fig. 1,  is strongly  $r$ dependent.
If $r$ is near 0, $E(x,t)$ is approxmately equal to $\cos [K(x-v_0t)]$.
If $r$ is near 1, Eq. (\ref{19}) shows an obvious shock-wave
structure. In some senses,  $r$ is related to  
the product $N_D^*\Lambda$ in
Kroemer's criterion for domain formation. To see this,
we first consider $r\rightarrow 1$. In this case, 
$(|\Psi_q|-|\Psi_{q+1}|)/|\Psi_q|$, and hence $(|a_2|-|a_1|)/|a_1|$,
is much smaller than 1 and there are
many excited unstable Fourier modes. Thus, $-eN_D^*v_0^{(1)}/\epsilon$
will be much larger than $D_nK^2$ and it follows from Eq. (\ref{16})
that Kroemer's criterion for sharp domain formation is 
$a_1 \Lambda /v_0 \approx eN_D^* v_0^{(1)} \Lambda /(v_0 \epsilon) \gg 1$, 
i.e. $N_D^*\Lambda\gg -v_0\epsilon e^{-1}/{v_0^{(1)}}$ ($\sim 10^{12}$
cm$^{-2}$). On the other hand, in the case $r\rightarrow 0$,
$-eN_D^*v_0^{(1)}/\epsilon$ is of the same order as $D_nK^2$ and
Kroemer's criterion for the appearance of sinusoidal electric-field 
domain is 
$a_1 \Lambda /v_0 \approx eN_D^* v_0^{(1)}\Lambda /(v_0 \epsilon) \approx 1$, 
i.e. $N_D^*\Lambda\approx -v_0\epsilon e^{-1}/{v_0^{(1)}}$ ($\sim 10^{12}$ cm$^{-2}$).
Therefore, Kroemer's criterion appears in Eq. (\ref{19}) via the 
representation of $r$ parameter. To determine the relation
between $r$ and $N_D^*\Lambda$, we can measure $(E_{max}-E_0)/(E_0-E_{min})$
($= (1+r)/(1-r)$) at different values of $N_D^*\Lambda$. Then we will
find that $r$ is a function of $N_D^*\Lambda$. In other words, $r$ can be 
determined by experiments. 

In order to get hole distribution, we assume 
$p(x,t)=
p_0+[p_s(x,t)e^{i\bar {K}(x+\mu_p E_0t)}
+p_s^*(x,t)e^{-i\bar {K}(x+\mu_p E_0t)}]/2$,
and substitute such $p(x,t)$ into Eq. (\ref{9}) to find 

\begin{equation}
\frac {\partial {p_s}}{\partial t}=-(D_p{\bar K}^2+\gamma N_D^*)p_s+
    D_p\frac {\partial^2 {p_s}}{\partial x^2}+
    (2i\bar {K}D_p+\mu_p E_0)\frac {\partial {p_s}}{\partial x}+
    S(x,t).
\label{21}
\end{equation}
Here $S(x,t)= 2gI_0e^{-i\bar {K}(x+\mu_p E_0 t)}m\cos (Kx+\Omega t)
+p_0(\mu_p-\gamma\epsilon/e)(\frac {\partial {E_s}}{\partial x}
+i\bar {K}E_s) e^{-i\bar {K}(v_0+\mu_p E_0)t}. $   
 Equation (\ref{21}) is a linear partial differential 
equation with external
driving function $S(x,t)$ including the MIP
and multiple Gunn domains. Thus $p_s$ can be considered as a
linear combination of a homogeneous solution $p_s^h(x,t)$ and
an inhomogeneous solution $p_s^i(x,t)$. 
Since the characteristic
length of $p_s$ is equal to $\Lambda$ ($\ll L$), we may treat
$\bar K$ as zero in Eq. (\ref{21})  and write  
$ {\partial {p_s^h}}/{\partial t}=-\gamma N_D^*p_s^h+
    D_p {\partial^2 {p_s^h}}/{\partial x^2}+
    \mu_p E_0 {\partial {p_s^h}}/{\partial x}$,
\begin{equation}
\frac {\partial {p_s^i}}{\partial t}=-\gamma N_D^*p_s^i+
    D_p\frac {\partial^2 {p_s^i}}{\partial x^2}+
    \mu_p E_0\frac {\partial {p_s^i}}{\partial x}+
    2gI_0m\cos (Kx+\Omega t)+
p_0(\mu_p-\gamma\frac {\epsilon}{e})\frac {\partial {E_s}}{\partial x}.
\label{23}
\end{equation} 
It is easy to find solution of $p_s^h$ and show that
$p_s^h=0$ as $t \to \infty$. 
When 
$E_s=\sum_{q=1}^{\infty}r^{q-1}(-a_1/b)^{1/2}$exp$[iqK(x-v_0t)]$ is
substituted into Eq. (\ref{23}), we can find solution for $p_s^i$.
Therefore, the hole distribution $p(x,t)=p_0+ Re(p_s^i)$ is
\begin{eqnarray}
&&p(x,t)=\frac {gI_0}{\gamma N_D^*}+\frac{2gI_0m}{A}
[(D_pK^2+\gamma N_D^*)\sin (Kx+\Omega t)-
(K\mu_p E_0-\Omega)\cos (Kx+\Omega t)]\nonumber\\
&&-D \sum_{q=1}^{\infty}\frac {qr^{q-1}\{q(v_0+\mu_p E_0)
\cos [qK(x-v_0t)]+(\gamma N_D^*+q^2K^2D_p)\sin [qK(x-v_0t)]\}}
{(\gamma N_D^*+q^2K^2D_p)^2+q^2K^2(v_0+\mu_p E_0)^2}. 
\label{26}
\end{eqnarray}
Here  $A=(D_pK^2+\gamma N_D^*)^2+(K\mu_p E_0-\Omega)^2$ and
$D=  {gI_0K}
(\mu_p-\gamma\epsilon/e)(-a_1/b)^{1/2}/{\gamma N_D^*}$.
The pattern profiles of Eqs. (\ref{19}) and (\ref{26})
correspond to simple dynamics.

When $I_0$ is large and the higher order corrections in 
Eqs. (\ref{2})-(\ref{5}) 
need to be included, then we have
\begin{equation}
{\partial E}/{\partial x}- e (n-N_D^*)/\epsilon=
-\bar\epsilon {e}/{\epsilon}p^{(0)},
\label{27}
\end{equation}
\begin{equation}
\frac {\partial n}{\partial t}+\frac {\partial}{\partial x}
\left[nv(E)-D_n\frac {\partial n}{\partial x}\right]=
\bar\epsilon\left[ gI(x,t)-\gamma (N_D^*+\frac {\epsilon}{e}
\frac{\partial E^{(0)}}{\partial x}) p^{(0)}\right],
\label{28}
\end{equation}
where $p^{(0)}$ and $E^{(0)}$ are  
solutions of Eqs. (\ref{26}) and (\ref{19}), 
respectively.
These results imply that up to order $O({\bar\epsilon})$ we have
\begin{equation}
\frac{\partial E}{\partial t}+\frac 1{\epsilon}eN_D^*{v}(E)
   +{v}(E)\frac{\partial E}{\partial x}-{D_n}\frac
{\partial^2E}{\partial x^2}-\frac 1{\epsilon}J_{tot}(t)=\bar S (x,t), 
\label{29}
\end{equation} 
where $\bar S (x,t)=
\bar\epsilon e \left[p^{(0)}v(E^{(0)})
-\mu_p E^{(0)}p^{(0)}-(D_p+D_n)
 {\partial p^{(0)}}/{\partial x}\right ]/\epsilon$
and represents the electron-hole coupling.
To understand the underlying physics of Eq. (\ref{29}),
we follow the same procedure as the case of simple dynamics,
to get 
\begin{equation}
\frac {d\Psi}{dt}= -C \Psi
      -\frac 18\left (\frac e\epsilon N_D^*v_0^{(3)}+iKv_0^{(2)}\right )
      {|\Psi|}^2\Psi+\int_0^{\Lambda} \bar S (x,t)e^{-iKx}dx,
\label{30}
\end{equation}
which is a generalization of Eq. (\ref{13}) and represents 
a self-sustained oscillator (i.e., PDT) under the external
driving force $\bar \epsilon F(t)/\epsilon$ 
($\equiv \int_0^{\Lambda} \bar S (x,t)e^{-iKx}dx$) and

\begin{eqnarray}
&F&(t)
=\left[iE_0\mu_p\frac \pi{K}-iv_0\frac \pi{K}-(D_p+D_n)\pi
\right ]G_1e^{i\Omega t}
-\left[E_0\mu_p\frac \pi{K}-v_0\frac \pi{K}+i(D_p+D_n)\pi
\right ]G_2e^{i\Omega t}\nonumber\\
&-&\left[E_0\mu_p\frac \pi{K}-v_0\frac \pi{K}+i(D_p+D_n)\pi
\right ]G_3e^{-iKv_0t}
+\left[iE_0\mu_p\frac \pi{K}-iv_0\frac \pi{K}-(D_p+D_n)\pi
\right ]G_4e^{-iKv_0t}\nonumber\\
&+&\frac \pi{K}\left(\frac {-a_1}b\right )^{1/2}
\left(v_0^{(1)}-\frac {a_1}{8b}v_0^{(3)}-\mu_p\right )
G_0e^{-iKv_0t},
\label{31}
\end{eqnarray}
where $ G_0= gI_0/\gamma N_D^*,\hspace{.05in}
G_1= 2gI_0m(D_pK^2+\gamma N_D^*)/A,\hspace{.05in}
G_2=-2gI_0m(K\mu_p E_0-\Omega)/A$,
$ G_3= -gI_0K^2 (\mu_p-\gamma\epsilon/e)(-a_1/b)^{1/2}
(v_0+\mu_p E_0)/B$,
$ G_4= -gI_0K (\mu_p-\gamma\epsilon/e)(-a_1/b)^{1/2}
(\gamma N_D^*+K^2D_p)/B$,
and $B=\gamma N_D^*(\gamma N_D^*+K^2D_p)^2+K^2(v_0+\mu_p E_0)^2$.
The intrinsic oscillating frequency in Eq. (\ref{30}) is $v_0/\Lambda$
and the external driving frequencies include $v_0/\Lambda$ and
$\Omega/2\pi$. 
If the operating point is in the PDM regime
(i.e., $a_1>0$), 
the last three terms in Eq. (\ref{31}) will be zero. This means that 
there is no Gunn-domain formation in the semiconductor,
but the external optical waves, i.e., the first two terms on the 
right-hand side of Eq. (\ref{31}),
will trigger a space-charge field (SCF)
$\sim e^{i(Kx+\Omega t)}$, which is well known in
photorefractive material \cite{ye}.
On the other hand, if the operating point is in the NDM regime,
in addition to the SCF, the PDT will be generated by the intrinsic
electrical instability with the 
oscillating frequency $Kv_0$. 
The relative strength of
the SCF and PDT depends on the strength of spatially independent part of
the MIP, i.e. $I_0$. When $I_0$ is small, SCF may be neglected
and the behavior of the system is dominated by the PDT.
When $I_0$ is large, the SCF will compete with the PDT and the system
will show complex dynamics. For example, these two competing frequencies
$\Omega/2\pi$ and $v_0/\Lambda$ can lead to quasiperiodic behavior. 

The picture presented above is useful for understanding
the results of other groups.
The system numerically studied by Suba\v cius {\em et al.} \cite{Sub}
may be represented by Eqs.(\ref{2})-(\ref{5}) of this Communication 
with $N^*_D=0$ and $\Omega =0$.
When the system is illuminated by a pulse of light interference
field with intensity $I(x,t)=I_0f(t)[1+m \cos(Kx)]$ and
$m <<1$ ($f(t)$ is the temporal shape of a laser pulse), 
the temporal part $I_0f(t)$ generates background
electrons in the conduction band and the very small spatially dependent
part of the light makes a small modulation on the background.
The situation is similar to the system of the present manuscript with
very small $I_0$. Fig. 1 in the present Communication is very
similar as the numerical results of Suba\v cius {\em et al.}.  
On the other hand, Segev {\em et al.} \cite{Seg} and 
Bonilla {\em et al.} \cite{Bon} considered deep-impurity doped 
GaAs under the externally optical waves, i.e.,
$I(x,t)=I_0[1+m \cos(Kx+\Omega t)]$.
In their case the modulation depth $m$ also plays a crucial role.
For example, Segev {\em et al.} considered $m$ as a small parameter.
Then the dynamical variables can be analytically expanded to Fourier series
based on the form of $I(x,t)$.
Therefore they can get multiple Gunn-domain order by order.  
Actually the physical idea of Segev {\em et al.} is the same as
Suba\v cius {\em et al.}. The spatial period of $I(x,t)$ 
determines the spacing of Gunn domains and $I_0$ (or $I_0f(t)$)
generates background electrons.
However the work of Segev {\em et al.} didn't discuss the stability of 
multiple Gunn-domain, which will lead to the travelling domains
with the same moving velocity (i.e., $\Omega/K$) of $I(x,t)$.
Furthermore, according to the prediction of Segev {\em et al.} the
moving velocity of electric-field profile in PDM or NDM regime  
has no any difference. This unsuitable prediction can be addressed 
and verified by Eq. ({\ref 30}). As for the numerical work of Bonilla {\em et al.},
they showed complicated spatiotemporal behaviors of Gunn domains, e.g.,
quasiperiodic route to chaos, are mostly observed in their simulations.
In other words, PDT didn't appear in their system. Therefore they concluded
that Kroemer's criterion can not give the correct numbers of Gunn domains.
We think that $m=0.1$ used in their study is large enough to
generate driven chaos due to the competition between SCF and PDT.
Therefore both considerations of Kroemer's criterion and pattern competition
may give correct physics in light-triggered Gunn-domain systems.

In conclusion, using a nonlinear analytic method we confirm the
LSA results of \cite{s1} and find that the competition
between the SCF and the PDT could be a dominant factor to determine the 
spatiotemporal pattern in semiconductor.   
We hope that the formulation of the present Communication 
can be further verified by numerical calculations or experiments.

We thank Jonathan Dushoff
for a critical reading of the paper.
This work was supported in part by the National Science Council of the
Republic of China (Taiwan) under Contract No. NSC 89-2112-M-001-084.

\newpage
{\bf{Figure Caption}}
\vskip 20 mm

\noindent  
 FIG. 1. The profile of $E(x,t)$ represented by Eq. (\ref{19}) for several
  values of $r$, where $E_0=10$ kV/cm, $(-a_1/b)^{1/2}=9.8$ kV/cm,
        $\Lambda =15$ $\mu$m, and $L=45$ $\mu$m.

\end{document}